\documentclass[12pt]{article}
\usepackage{amsfonts}

\usepackage{amsfonts}
\usepackage{amssymb}
\usepackage{amscd}

\date{}

\begin{document}

 \title{\Large\bf A $p$-Adic Model \\ of DNA Sequence and Genetic Code}

\author{Branko
Dragovich$^a$\,\footnote{\textsf{\, E-mail:\,dragovich@phy.bg.ac.yu}} and Alexandra Dragovich$^b$\\
 {\it $^a$Institute of Physics, P.O. Box 57,} \\ {\it 11001
Belgrade, Serbia} \\ {\it $^b$Vavilov Institute of General
Genetics}\\ {\it Gubkin St. 3, \, 119991  Moscow, \,Russia}}

\maketitle

\begin{abstract}
Using basic properties of $p$-adic numbers, we consider a simple new
approach to describe main aspects of DNA sequence and genetic code.
Central role in our investigation plays an ultrametric $p$-adic
information space which basic elements are nucleotides, codons and
genes. We show that a $5$-adic model is appropriate for DNA
sequence. This $5$-adic model, combined with $2$-adic distance, is
also suitable for genetic code and for a more advanced employment in
genomics. We find that genetic code degeneracy is related to the
$p$-adic distance between codons.
\end{abstract}

\bigskip

\section{Introduction}

It is well known that practically all genetic information in living
systems is contained in the desoxyribonucleic acid (DNA) sequence.
The DNA macromolecules are made of two polynucleotide chains  with a
double-helical structure. There are four nucleotides called: adenine
(A), guanine (G), cytosine (C) and thymine (T). A and G belong to
purine, while C and T to pyrimidine. The DNA is packaged into
chromosome which is localized in the nucleus of the eukaryotic
cells. One of the basic processes within DNA is its replication. The
passage of its gene information to protein, called gene expression,
performs by the messenger ribonucleic acid (mRNA), which is usually
a single polynucleotide chain. In the first part of this process,
known as transcription, the nucleotides A, G, C, T from DNA are
respectively transcribed into the nucleotides U, C, G, A of mRNA,
i.e. T is replaced by U, where U is the uracil. The next step is
translation, when  mRNA codon information is translated into
synthesis of proteins. Codons are ordered sequences of three
nucleotides of the A, G, C, U. Protein synthesis in all eukaryotic
cells performs in the cytoplasm. The genes by their codons control
amino-acid sequences in proteins. It is obvious that there are $4
\times 4\times 4 = 64$ possible codons. However $61$ of them specify
the $20$ different amino-acids and $3$ correspond to stop-codons,
which serve as termination signals. As a result most amino-acids are
encoded by more than one codon. This degenerate correspondence
between codons and amino-acids is known as genetic code, which is
mostly universal for all living organisms. In almost all cells
genetic information flows from DNA to RNA to protein. For a detail
and comprehensive information on molecular biology aspects of DNA,
RNA and genetic code one can see Ref. \cite{watson}.

Processes within macromolecules can be regarded as quantum as
classical depending on the scale we are interested in. Modeling of
DNA, RNA and genetic code is a challenge as well as a chance for
modern mathematical physics. An interesting model based on the
quantum algebra $\mathcal{U}_q (sl(2)\oplus sl(2))$ in the $q \to 0$
limit was proposed as a symmetry algebra for the genetic code (see
\cite{sorba}, \cite{sorba1} and references therein). In a sense this
approach mimics quark model of baryons. To describe correspondence
between codons  and amino-acids, it was constructed an operator
which acts on the space of codons and its eigenvalues are related to
amino-acids. Besides some successes of this approach, there is a
problem with rather many parameters in the operator.

There are some very complex systems (e.g. spin glasses and some
macromolecules) whose space of states has an ultrametric structure.
The space of conformational states of proteins is such one.
Processes on ultrametric spaces usually need new methods for their
description. $p$-Adic models with pseudodifferential operators have
been successfully  applied to interbasin kinetics of proteins
\cite{avetisov1}, \cite{avetisov2}, \cite{avetisov3} (for a brief
review see \cite{kozyrev}). Ultrametricity is a suitable
mathematical concept and a tool for description of systems with
hierarchical structure. The first field of science where
ultrametricity observed was taxonomy. The first review of
ultrametricity in physics and biology was presented twenty years ago
\cite{rammal}. A very significant and promising part of ultrametrics
is p-adics.

 $p$-Adic numbers are discovered at the end of the 19th century by
 German  mathematician Kurt Hensel.  They have been successfully
 employed in many parts of mathematics. Since 1987 they have been also used
 in construction of various physical models, especially in string
 theory, quantum mechanics, quantum cosmology and dynamical systems (for a
review, see \cite{freund} and
 \cite{vladimirov1}).  Some $p$-adic aspects of cognitive, psychological and social phenomena
 have been also considered \cite{khrennikov1}. The present status of application of
 $p$-adic numbers in physics and related branches of sciences is
 reflected in the proceedings of the 2nd International Conference on
 $p$-Adic Mathematical Physics \cite{proceedings}.

 A $p$-adic approach to
genetics has not been tempted so far. The main aim of this paper is
to make the first step towards $p$-adic genomics. Starting with a
formulation of $p$-adic genetic information space, we propose a
$5$-adic model for DNA (and RNA) sequences and genetic code. A
central mathematical tool to analyze  classification of codons and
structure of genetic code is $p$-adic distance between codons.

\section{$p$-Adic numbers}

Recall that numerical results of measurements in experiments  and
observations are rational numbers. The set of all rational numbers
$\mathbb{Q}$, having usual properties of summation and
multiplication, is algebraically a field. In addition to arithmetic
operations it is often important to know also a distance between
numbers. Distance can be defined by a norm. On $\mathbb{Q}$ there
are two kinds of nontrivial norm: usual absolute value
$|\cdot|_\infty$ and $p$-adic absolute value $|\cdot|_p$ , where $p$
is any prime number. The usual absolute value is well known from
elementary courses of mathematics and the corresponding distance
between two real numbers $x$ and $y$ is $d_\infty (x, y) =
|x-y|_\infty$. This distance also enables that all infinite decimal
expansions of real numbers
\begin{equation}
 x = \pm \, 10^n \sum_{k=0}^{-\infty} a_k \, 10^k \,, \quad a_k \in \{ 0,\, 1,
 \cdots , 9 \}\,,  \,\,\, a_0 \neq 0 \,, \, n\in \mathbb{Z}  \label{2.1}
\end{equation}
are convergent.

By definition, $p$-adic norm of a rational number $0 \neq x = p^\nu
\, \frac{r}{s}$, where $ \nu \in \mathbb{Z}$, and integers $r$ and
$s$ are not divisible by given prime number $p$, is $|x|_p =
p^{-\nu}$, and $|0|_p =0.$ This norm is a mapping from $\mathbb{Q}$
into non-negative real numbers and has the following properties:

(i) $|x|_p \geq 0, \,\, |x|_p =0$ if and only if $x = 0$,

(ii) $|x\, y|_p = |x|_p \,  |y|_p \,,$

(iii) $|x + y|_p \leq \, max\, \{ |x|_p\,, |y|_p \} \leq |x|_p +
|y|_p $ for all $x \,, y \in \mathbb{Q}$.

\noindent Because of the strong triangle inequality $|x + y|_p \leq
\, max \{ |x|_p\,, |y|_p \}$ $p$-adic absolute value belongs to
non-Archimedean (or ultrametric) norm.

$p$-Adic distance between two rational numbers $x$ and $y$ is
\begin{equation}
d_p (x\,, y) = |x - y|_p \,.    \label{2.2}
\end{equation}
Since $p$-adic absolute value is ultrametric, the $p$-adic distance
(\ref{2.2}) is also ultrametric, i.e. it satisfies
\begin{equation}
d_p (x\,, y) \leq\, max\, \{ d_p (x\,, z) \,, d_p (z\,, y) \} \leq
d_p (x\,, z) + d_p (z\,, y) \,, \label{2.3}
\end{equation}
where $x, \, y$ and $z$ are any three points of a $p$-adic space.

In direct analogy with the field $\mathbb{R}$ of real numbers,  the
field $\mathbb{Q}_p$ of $p$-adic numbers can be introduced by
completion of $\mathbb{Q}$ with respect to the distance (\ref{2.2}).
Note tat for each prime $p$ there is one $\mathbb{Q}_p$. Any $x \in
\mathbb{Q}_p$ has a unique expansion
\begin{equation}
 x = p^m \sum_{k=0}^{+\infty} a_k \, p^k \,, \quad a_k \in \{ 0,\, 1,
 \cdots , p-1 \}\,,  \,\,\, a_0 \neq 0 \,,  \label{2.4}
\end{equation}
where $m$ is an ordinary integer.

In this paper we use only $p$-adic integers for which $m = 0,\, 1,\,
2,\, \cdots $.

For a simple introduction into $p$-adic numbers one can see book
\cite{gouvea}.

\section{$p$-Adic Genetic Information Space}

We want to present now a mathematical formalism suitable for
modeling  genetic code and DNA sequence. Let us first introduce an
{\it information space} $\mathcal{I}$ as a subset of the set
$\mathbb{Z}$ of usual integer numbers, where to each $m \in
\mathcal{I}$ is attached an information. Different numbers $a , b
\in \mathcal{I}$ contain different information. Let be valid
standard arithmetic operations (summation, subtraction and
multiplication) on elements of $\mathcal{I}$.

 Since an information can be more or less similar (or
dissimilar) to another, there is a sense to introduce a mathematical
tool  to measure  similarity (or dissimilarity). Such a tool is a
distance between the corresponding integers. But now arises a
question: What kind of distance we should take between integers to
describe closeness on the information space? Recall that there are
two kinds of distances for integers: usual real (Archimedean) and
$p$-adic (non-Archimedean, ultrametric) distance. We propose, for a
class of $\mathcal{I}$, to employ $p$-adic distance (defined in the
preceding section), i.e. $d_p (a, b) = |a -b|_p \,, \quad a, b \in
\mathbb{Z}$.  As a consequence one has a quite natural property: two
information are   closer, i.e. with smaller distance, if they have
more equal first digits in their $p$-adic expansion. One has also
that digits which come later in the expansion have smaller
importance (for a similar treatment of information see
\cite{khrennikov2}). In the sequel an information space with
$p$-adic distance will be called {\it $p$-adic information space}.
Some experimental properties of genetic code lead us to introduce
{\it $p$-adic genetic space} $\mathcal{G}_p$ as a  special case of
$p$-adic $\mathcal{I}$. An element $m\in \mathcal{G}_p$ can be
presented in the form
\begin{equation}
m = \pm \, p^N \, \sum_{i=0}^{n}  m_i \, p^i   \,, \quad m_i \in \{
0\,, 1\,, \cdots , p-1 \}\,, \label{3.5}
\end{equation}
where $ N \,, n$ are nonnegative integers and $m_i$ are digits. For
a given $p$ and $N$, information $m$ is characterized by the
sequence of digits $m_0\,, m_1 \,, \cdots \,, m_n$. In other words,
information is coded by  ordered sequence of digits $m_0\,, m_1 \,,
\cdots \,, m_n$. If integers $a\,, b\in \mathcal{G}_p$ have
expansions
\begin{equation}
a = a_0 + a_1\, p + a_2\, p^2 + \cdots \,, \quad \quad  b = b_0 +
b_1\, p + b_2\, p^2 + \cdots \,,
\end{equation}
then $d_p (a, b) = p^{-k}$ if $a_0 = b_0 \,, \cdots \,, a_{k-1} =
b_{k-1} $ and $a_k \neq b_k$. Accordingly $d_p (a, b) = p^{-k}$ is
smaller as $k$ is larger and $a\,, b$ are closer (i.e. more
similar). This $p$-adic closeness will be later exploited in
analysis of genetic code degeneration, but now let us turn to the
$p$-adic modeling of DNA.

\section{$p$-Adic model of the DNA sequence}

To have an appropriate $p$-adic genetic space $\mathcal{G}_p$ that
can describe DNA sequence and genetic code, one has to choose the
corresponding prime number $p$ which will be used as a base for
expansion. For the base in expansion of genetic information we
choose $ p=5 $, because $5$ is the smallest prime number which
contains four nucleotides (A\,, T\,, G\,, C) in DNA, or (A\,, U\,,
G\,, C) in RNA, in the form of four different digits. At the first
glance, because there are four nucleotides, one could start to think
that a $4$-adic expansion, which has just four digits, might be more
appropriate. However, note that $4$ is a composite integer and that
related expansion is not suitable since the corresponding
$|\cdot|_4$ absolute value is not a norm but a pseudonorm and it
makes a problem with uniqueness of the distance between two points.
To illustrate this problem let us consider, for instance, a distance
between numbers $4$ and $0$. Then we have $d_4 (0,4) = |4|_4 =
\frac{1}{4}$, but on the other hand $d_4 (0,4) = |2|_4 \, |2|_4 =
1$.

Thus for four nucleotides, which appear in the strict
complementarity  between the two DNA strands, i.e. make two base
pairs $(A, T)$ and $(C, G)$, we choose the corresponding $5$-adic
integer numbers to construct the corresponding DNA sequence model.
Namely, we attach digits $(1,\, 2,\, 3,\, 4)$ to nucleotides $(C,\,
A,\, T,\, G)$ in the following way:
\begin{equation} C= 1,\,\, A=2,\,\, T= 3,\,\, G=4
\,. \label{4.1} \end{equation} Recall that there are $p$ digits in
representation of a $p$-adic number. According to this approach, the
digit $0$ does not play a role in the representation of single
helicoidal chain and in the RNA coding. It is worth noting that we
also considered some other choices of possible connection between
nucleotides and four of the above five digits. However, we find that
the choice (\ref{4.1}) is the most suitable and attractive.

 In this way any of
the DNA chains can be presented as a $5$-adic number  in the form
\begin{equation} x =5^N (x_0 + x_1\, 5 + x_2 \, 5^2 + \cdots + x_n \, 5^n )  \,,
\quad x_i \neq 0   \,, \quad N \in \mathbb{N} \cup \{ 0\}\,,\,\,n\in
\mathbb{N} \,,\label{3.1}
\end{equation}
where $x_i = 1,\, 2,\, 3,\, 4$ and $n$ is an enough large natural
number. This chain can be also presented as
\begin{equation}
x =\sum_{j=1}^\omega 5^{N_j} (x_0 + x_1\, 5 + x_2 \, 5^2 + \cdots +
x_{n_j} \, 5^{n_j} ) \,, \quad N_1 < N_2 < \cdots N_\omega \,,
\end{equation}
where $\omega$ is a number of subsequences, which encode and those
which do not encode proteins,  in a chain of the DNA. One can
introduce $5$-adic distance between genes and it will be
characterized by $5^{-N_j}$.

 For a simple illustrative example $(N = 0,\, n=10)$, to a chain of nucleotides
\begin{equation}
a = A\, T\, G\, C\, A\, A\, G\, T\, G\, A       \label{3.2}
\end{equation}
corresponds $5$-adic number
\begin{equation}
a = 2 + 3\cdot 5 + 4\cdot 5^2 + 1\cdot 5^3 +  2\cdot 5^4 + 2\cdot
5^5 + 4\cdot 5^6 + 3\cdot 5^7 + 4\cdot 5^8 + 2\cdot 5^9 \,,
\label{3.3}
\end{equation}
which can be written also using only its digits
\begin{equation}
a =2\, 3\, 4\, 1\, 2\, 2\,  4\, 3\, 4\, 2 \,.       \label{3.4}
\end{equation}
According to this approach a DNA sequence can be presented as a sum
of two $5$-adic integers. Let us denote DNA sequences  by Greek
letters $\alpha, \, \beta, \cdots$ and their chain components by
Latin ones $a,\, b ,\cdots$. Then an $\alpha = a + b$. In fact $a$
and $b$ are firmly correlated because of complementarity, i.e. $b =
\bar{a}$, where $\bar{a}$ obtains from $a$ replacing digits $(1,\,
2,\, 3\,, 4)$ by $(4,\, 3,\, 2\,, 1)$, respectively. The
corresponding $\alpha$ related to (\ref{3.2}) is
\begin{equation}
\alpha = a + \bar{a} = 2\, 3\, 4\, 1\, 2\, 2\,  4\, 3\, 4\, 2 \, +
\, 3\, 2\, 1\, 4\, 3\, 3\,  1\, 2\, 1\, 3 \, = \, 0\, 1\, 1\, 1\,
1\, 1\, 1\, 1\, 1\, 1\, 1 , \label{3.3}
\end{equation}
where we performed summation of digits from the left to the right,
taking $1 + 4 = 0 + 1\cdot 5$ and $2+ 3 = 0 + 1 \cdot 5$.  In this
way the sum (\ref{3.3}), which corresponds to an example of DNA, is
presented in the very simple form: it is quite definite sequence of
the digit $1$, which is of the same length as DNA and shifted at one
place on the right.

One can easily check that integers $a\,, \bar{a}$ and $\alpha$ in
(\ref{3.3}) form vertices of an equilateral triangle whose all three
sides have the same $5$-adic length equal to $1$.

It is worth mentioning that human genome, which presents all genetic
information of the organism, is composed of more than three billion
base pairs and contains more than 30.000 genes.

\section{$p$-Adic genetic code}

A living cell is a very complex system composed mainly of protein
macromolecules  playing various roles. All those proteins are made
of only 20 amino-acids, which are the same for all living world on
the Earth. Different sequences of amino-acids form different
proteins. An intensive study of connection between ordering of
nucleotides in the DNA (and RNA) and ordering of amino-acids in
proteins led to the discovery of genetic code.

At the end of the 50th and beginning of the 60th of the last century
many basic properties of genetic code were obtained. Genetic code is
understood as a dictionary for translation of information from the
DNA (through RNA) to production of proteins by amino-acids. The
information is contained in codons, which are ordered sequences of
three nucleotides. There are three stop codons, and 61 codons are
related to 20 amino-acids. There are  various multiplicity (one,
two, three, four and six) of codons which correspond to amino-acids
in proteins, i.e. genetic code is degenerate. This is an well
established experimental fact.

However, there is no simple theoretical understanding of genetic
coding. In particular, it is not clear why genetic code is just in
the known way and not in many other possible ways. What is a
principle (or principles) used in fixing  mitochondrial and
eukaryotic codes? What are properties of codons responsible for
their appearance in quadruplets, sextets, doublets, and even in a
triplet and a singlet. These are only some of many questions which
can be asked about genetic code. Recall that the ribosome performs
synthesis of proteins and it knows somehow very firmly which
amino-acid corresponds to a given codon. In fact, the ribosome is a
molecular machine which performs multiple functions, and one of them
should be a computing of codon properties.

Let us consider now possible answers to the above questions on
genetic code starting from the $5$-adic model. According to our
approach, a codon in RNA is an integer number of the following form:
\begin{equation}
c = c_0 + c_1 \, 5 + c_2 \, 5^2 \,, \quad c_0\,, c_1\,, c_2 \in \{
1\,, 2\,, 3\,, 4\} \,,  \label{5.1}
\end{equation}
where, without loss of generality, we take $N =0$.
 In the RNA the nucleotide T
is replaced by U and we  remain the same digit (T=3) and take  U=3.
In this way there is no digit $0$ used in presentation of codons.

Having  the above choice  of digits (i.e. C=1,\, A=2,\, U=3,\, G=4)
we can now look at the Tables 1 and  2, and  observe the
corresponding ultrametric ($5$-adic and $2$-adic) reason for
formation of quadruplets and doublets. Codons are simultaneously
denoted by three digits and capital letters. The corresponding
amino-acids are presented in the usual three letters form.

\noindent 
\begin{tabular}{|c|c|c|c|}
 \hline \ & \ & \ & \\
 111 CCC Pro &  211 ACC Thr  & 311 UCC Ser & 411 GCC Ala  \\
 112 CCA Pro &  212 ACA Thr  & 312 UCA Ser & 412 GCA Ala  \\
 113 CCU Pro &  213 ACU Thr  & 313 UCU Ser & 413 GCU Ala  \\
 114 CCG Pro &  214 ACG Thr  & 314 UCG Ser & 414 GCG Ala  \\
 \hline \  & \  &  \ & \ \\
 121 CAC His &  221 AAC Asn  & 321 UAC Tyr & 421 GAC Asp  \\
 122 CAA Gln &  222 UAA Lys  & 322 UAA Ter & 422 GAA Glu  \\
 123 CAU His &  223 AAU Asn  & 323 UAU Tyr & 423 GAU Asp  \\
 124 CAG Gln &  224 AAG Lys  & 324 UAG Ter & 424 GAG Glu  \\
 \hline \  & \  & \  &   \\
 131 CUC Leu &  231 AUC Ile  & 331 UUC Phe & 431 GUC Val \\
 132 CUA Leu &  232 AUA Met  & 332 UUA Leu & 432 GUA Val \\
 133 CUU Leu &  233 AUU Ile  & 333 UUU Phe & 433 GUU Val \\
 134 CUG Leu &  234 AUG Met  & 334 UUG Leu & 434 GUG Val \\
 \hline \ & \   & \  &   \\
 141 CGC Arg &  241 AGC Ser  & 341 UGC Cys & 441 GGC Gly  \\
 142 CGA Arg &  242 AGA Ter  & 342 UGA Trp & 442 GGA Gly  \\
 143 CGU Arg &  243 AGU Ser  & 343 UGU Cys & 443 GGU Gly  \\
 144 CGG Arg &  244 AGG Ter  & 344 UGG Trp & 444 GGG Gly  \\
\hline
\end{tabular}

\bigskip
\centerline{Table 1 : The vertebral mitochondrial code}
\bigskip

\noindent 
\begin{tabular}{|c|c|c|c|}
 \hline \ & \ & \ & \\
 111 CCC Pro &  211 ACC Thr  & 311 UCC Ser & 411 GCC Ala  \\
 112 CCA Pro &  212 ACA Thr  & 312 UCA Ser & 412 GCA Ala  \\
 113 CCU Pro &  213 ACU Thr  & 313 UCU Ser & 413 GCU Ala  \\
 114 CCG Pro &  214 ACG Thr  & 314 UCG Ser & 414 GCG Ala  \\
 \hline \  & \  &  \ & \ \\
 121 CAC His &  221 AAC Asn  & 321 UAC Tyr & 421 GAC Asp  \\
 122 CAA Gln &  222 UAA Lys  & 322 UAA Ter & 422 GAA Glu  \\
 123 CAU His &  223 AAU Asn  & 323 UAU Tyr & 423 GAU Asp  \\
 124 CAG Gln &  224 AAG Lys  & 324 UAG Ter & 424 GAG Glu  \\
 \hline \  & \  & \  &   \\
 131 CUC Leu &  231 AUC Ile  & 331 UUC Phe & 431 GUC Val \\
 132 CUA Leu &  232 AUA Ile  & 332 UUA Leu & 432 GUA Val \\
 133 CUU Leu &  233 AUU Ile  & 333 UUU Phe & 433 GUU Val \\
 134 CUG Leu &  234 AUG Met  & 334 UUG Leu & 434 GUG Val \\
 \hline \ & \   & \  &   \\
 141 CGC Arg &  241 AGC Ser  & 341 UGC Cys & 441 GGC Gly  \\
 142 CGA Arg &  242 AGA Arg  & 342 UGA Ter & 442 GGA Gly  \\
 143 CGU Arg &  243 AGU Ser  & 343 UGU Cys & 443 GGU Gly  \\
 144 CGG Arg &  244 AGG Arg  & 344 UGG Trp & 444 GGG Gly  \\
\hline
\end{tabular}

\bigskip
\centerline{Table 2 : The eucaryotic code}
\bigskip
\bigskip
Our observations are as follows.

(i) Codons with the same first two digits have the same $5$-adic
distance equal to $\frac{1}{25}$. This property leads to clustering
of $64$ codons into their $16$ quadruplets. Namely, any two codons
$a$ and $b$ whose the  first two digits are mutually equal and the
third one is different, have $5$-adic distance
\begin{equation}
d_5 (a,\, b) = |a_0 + a_1 \, 5 + a_2 \, 5^2 - (a_0 + a_1 \, 5 + b_2
\, 5^2)|_5 = |(a_2 - b_2) \, 5^2| = 5^{-2}\,,  \label{5.2}
\end{equation}
where $a_0 \,, a_1\,, a_2\,, b_2 \in \{1\,, 2\,, 3\,, 4  \}$ and
$a_2 \neq b_2$. Since $a_0$ and $a_1$ may have four values, there
are $16$ quadruplets.

 (ii) With respect to $2$-adic distance, the above clusters may be regarded
 as composed of two doublets: $a = a_0\, a_1\, 1$ and $b = a_0\, a_1\, 3$
 make the first doublet, and
 $c = a_0\, a_1\, 2$ and $d = a_0\, a_1\, 4$ form the second one. $2$-Adic
 distance between codons within each of these doublets is
 $\frac{1}{2}$, i.e.
 \begin{equation}
d_2 (a,\, b) = |(3 -1)\, 5^2|_2 =\frac{1}{2} \,, \quad  d_2 (c,\, d)
= |(4 -2)\, 5^2|_2 =\frac{1}{2} \,.   \label{5.3}
 \end{equation}

(iii) Quadruplets which have at the second position digit $1$ do
   not decay into two doublets. Each of these four quadruplets
   corresponds to the one of four different amino-acids.

(iv)  Quadruplets which have at the second position digit $2$ decay
into two doublets mentioned in (ii). Each of these eight doublets
corresponds to the one of the new eighth different amino-acids.

(v) The doublet structure of quadruplets which have at the second
position digit $3$ or $4$ becomes more complex and depend also on
digit at the first place. Quadruplets with digits $1\, 3\, i\,,
\quad 4\, 3\, i\,, \quad 1\, 4\, i$ and $4\, 4\, i$ , where $i \in
\{1\,, 2\,, 3\,, 4 \}$, are stable and have not substructure.
However, for other four combinations od the first two digits the
situation depends on the kind (mitochondrial or eukaryotic) of
coding. The situation is simple for the vertebral mitochondrial
code: quadruplets with digits $2\, 3\, i\,, \quad 3\, 3\, i\,, \quad
2\, 4\, i$ and $4\, 4\, i$ , where $i \in \{1\,, 2\,, 3\,, 4 \}$,
are not stable and decay into doublets. In the case of the
eukaryotic (universal) code one has: quadruplet with digits $2\, 3\,
i$ decays into one Ile-triplet $(2\, 3\, 1 \,, \,\, 2\, 3\, 2\,,
\,\, 2\, 3\, 3 )$ and one Met-singlet $2\, 3\, 4$, while the
quadruplet $3\, 4\, i$ separates into one doublet  and two different
singlets.

We would like to emphasize that codons ending on digits $1$ and $3$,
and having $2$-adic distance $\frac{1}{2}$, appear always together
and determine the same amino-acid.

\section{Concluding remarks}

In this paper we proposed a new and  simple model to investigate
information aspects of DNA, RNA and genetic code. To this end, we
introduced the corresponding $p$-adic information space and
connected it with DNA when $p = 5$.

An essential property of any $p$-adic space is ultrametric behavior
of distances between its elements, which radically differs from
usual distances on a space of real numbers. It is significant that
we attached just $5$-adic integer numbers to the sequence of codons
and not real integers in base $5$.

Classification of any set of objects is an ordering them into groups
according to some their relations. Using $5$-adic and $2$-adic
distances between codons we obtained their classification into
quadruplets and doublets, respectively. As a result of the above
analysis one obtains the following principle of genetic coding: {\it
$p$-adically close codons correspond to the same amino-acid.}

We plan to continue  research on this model and  to develop its
formalism as well as to apply it to more concrete cases.

\bigskip

\noindent{\bf Acknowledgments}

The work on this article was partially supported by the Ministry of
Science and Environmental Protection, Serbia, under contract No
144032D. One of the authors (B.D) would like to thank P. Sorba for
useful discussion and reading  a draft of this paper. He also thanks
V.S. Vladimirov, I.V. Volovich, A.Yu. Khrennikov and S.V. Kozyrev
for various valuable and inspiring  discussions on $p$-adics.

\end{document}